# Two-sided confidence interval of a binomial proportion: how to choose?

André GILLIBERT[ab]*[†], Jacques BÉNICHOU[bc] and Bruno FALISSARD[a]

[a] INSERM UMR 1178, Université Paris Sud, Maison de Solenn, Paris, France.

[b] Department of Biostatistics and Clinical Research, Rouen University Hospital, Rouen, France

[c] Inserm U 1219, Normandie University, Rouen, France

* Correspondence to: André GILLIBERT, Department of Biostatistics and Clinical Research, Rouen University Hospital, Rouen, France

[†]E-mail: andre.gillibert@chu-rouen.fr

# Abstract

**Introduction**: estimation of confidence intervals (CIs) of binomial proportions has been reviewed more than once but the directional interpretation, distinguishing the overestimation from the underestimation, was neglected while the sample size and theoretical proportion variances from experiment to experiment have not been formally taken in account. Herein, we define and apply new evaluation criteria, then give recommendations for the practical use of these CIs.

**Materials & methods**: Google® Scholar was used for bibliographic research. Evaluation criteria were (i) one-sided conditional errors, (ii) one-sided local average errors assuming a random theoretical proportion and (iii) expected half-widths of CIs.

**Results**: Wald's CI did not control any of the risks, even when the expected number of successes reached 32. The likelihood ratio CI had a better balance than the logistic Wald CI. The Clopper-Pearson mid-P CI controlled well one-sided local average errors whereas the simple Clopper-Pearson CI was strictly conservative on both one-sided conditional errors. The percentile and basic bootstrap CIs had the same bias order as Wald's CI whereas the studentized CIs and $BC_a$, modified for discrete bootstrap distributions, were less biased but not as efficient as the parametric methods. The half-widths of CIs mirrored local average errors.

**Conclusion**: we recommend using the Clopper-Pearson mid-P CI for the estimation of a proportion except for observed-theoretical proportion comparison under controlled experimental conditions in which the Clopper-Pearson CI may be better.

**KEYWORDS**: binomial confidence interval, coverage bias, equal-tailed confidence intervals, local average errors, Clopper-Pearson mid-P confidence interval.

# 1. Introduction

Estimating the confidence interval (CI) of a proportion is one of the most basic statistical problems and an everyday task for many statisticians. Most of statistical software provides two estimators, an "approximate" and an "exact" estimator. The "approximate" CI estimator is usually Wald's CI, namely

$$\hat{p} \pm z_{1-\alpha/2}\sqrt{\frac{\hat{p}(1-\hat{p})}{n}}, \qquad (1)$$

where $\hat{p}$ is the observed proportion and $n$ is the sample size. Wald's estimator may be the best known, but has been criticized for its coverage bias deemed unacceptable [1,2]. The "exact" CI estimator is usually the Clopper-Pearson CI [3] (*e.g.*, in R, SAS, Stata). The word "exact" is misleading since no deterministic estimator has complete control over the coverage due to the binomial distribution discreteness.

The unsolvable problem of exact coverage led to the development of many estimators: Newcombe reviewed seven estimators in 1998, then Pires reviewed as many as 20 different estimators in 2008 [4] and the development of new methods was still active as of 2018 [5]. Agresti and Coull [1] argue that the Clopper-Pearson estimator is too conservative, with actual coverage always greater or equal to the nominal coverage (strict conservatism) and suggest that some approximate estimators (e.g. Wilson and Agresti-Coull) may have better control over the average coverage than an exact estimator.

Some standard criteria for assessing the validity of CI estimators have been widely used in systematic reviews, such as minimal coverage control, average coverage control and interval width [1,2,4,6–8]. In our opinion, two important issues have not been sufficiently addressed.

The first issue is the balance of one-sided errors: equal-tailed CIs or unequal-tailed CIs. This is needed for a directional interpretation of CIs where overestimation and underestimation are distinguished [9]. For instance a 95% unequal-tailed CI estimator, applied to the rate of adverse events of a treatment, has an unpredictable probability between 0 and 5% of underestimating the proportion of adverse events, another unpredictable probability between 5% and 0% of overestimating the proportion with a sum of the two probabilities equal or close to 5%. In order to prove the safety of the treatment, the actual proportion could be claimed to be less than the upper boundary of the CI of the rate of adverse events; doing so, the actual risk taken is unpredictable, between 0% and 5%. An equal-tailed 95% CI estimator controls the probability of overestimating the actual proportion (equal or close to 2.5%) and the probability of underestimating it (equal or close to 2.5%). The actual risk taken when claiming that the rate of adverse events is less than the upper boundary would be controlled (close to 2.5%). In our opinion, directional interpretation is important and equal-tailed CIs should be preferred.

The second issue is the variability of the actual proportion and sample size from one experiment to another. Although most CI estimators are conditional to the sample size, this sample size is not actually constant from one experiment to another. Most meta-analyses show different sample sizes for all included research studies. Moreover, they show some heterogeneity of actual proportions or effects. Consequently, the actual CI coverage biases should be judged in a realistic setting where experiment replication is performed with a variable sample size, a variable proportion or both. This variability should smooth coverage oscillations described by Brown, Cai and DasGupta, attributed to discreteness of the binomial distribution [2].

The objective of this article is to systematically review binomial proportion and Poisson CI estimators and assess them with evaluation criteria taking in account the two issues mentioned above: directional interpretation and variability of the actual proportion or sample size.

# 2. Methods

## 2.1. Systematic review

A bibliographic search with the keywords "binomial", "confidence" and "interval" was conducted on the Google® Scholar database in July 2017 (updated in October 2018), looking for articles defining CIs of a binomial proportion. Articles were sorted by relevance according to Google's algorithm and the first 400 results were screened on their title then their summary. The references of systematic review articles were used to identify original references. The CI estimators considered as redundant will be mentioned but not presented. Estimators were considered redundant when they were very accurate approximations of other estimators and were graphically indistinguishable on error and width figures. Deterministic behavior and simple interpretation was considered a requirement. Consequently, randomized or fuzzy CIs have not been implemented. Even though some randomness is included in bootstrap CIs, they have been considered as deterministic since stability is achieved with a high number of bootstrap samples. The exact asymptotic solution to bootstrapping has been found, so bootstrap CIs presented in Appendix 1 are perfectly deterministic.

## 2.2. Evaluation criteria

### 2.2.1. One-sided conditional errors

Conditional errors are defined as coverage errors conditional to a constant sample size and constant actual proportion $p$. The experiment is a draw from a random binomial distribution $\mathcal{B}(n;p)$. The conditional lower bound error is noted $\alpha'_L$ and is defined as the actual probability that the CI is below the actual proportion $p$. Similarly, we denote $\alpha'_U$, the actual probability that the CI is above the actual proportion $p$.

### 2.2.2. One-sided local average errors

We define a random two-steps random binomial experiment. The first step is to draw an actual proportion $p$ from a random variable $P$, taking in account the variability of the actual proportion from experiment to experiment. The second step is to draw an actual binomial variable $x$ from a sample of size $n$ following a binomial distribution $\mathcal{B}(n;p)$. Then, a CI is computed from $x$ and $n$. The local average coverage is defined as the probability that the CI contains the realization $p$ of $P$.

The random variable $P$ will be modeled as a logit-normal variable such that the typical odds ratio of the actual proportion between two experiments is $OR_S$. That is, $\log(P) \sim \mathcal{N}(\mu; \sigma^2)$ and $\exp(\sigma) = OR_S$. For the primary analysis, $OR_S$ will be set at 1.20 (arbitrary choice). Sensitivity analyses with $OR_S = 1.10$ and $OR_S = 1.05$ will be performed. We will denote $p_0 = E[P]$ the expected proportion and $\lambda = np_0$ the expected number of events. The $\alpha_L''$ risk is the actual probability that the CI is below the realization of $P$ while $\alpha_U''$ is the actual probability that the CI is above the realization of $P$.

A sensitivity analysis will be performed with a random sample size $N$ and a constant proportion $p$ rather than a random proportion $P$ and a constant sample size $n$. It is hypothesized that both models will equally smooth error curves.

The Poisson distribution will be seen as the asymptotic case of the binomial distribution when the sample size tends towards positive infinity.

### 2.2.3. Local average half-widths

For a number of successes $x$ and a number of trials $n$, the lower and upper half-widths of the CI $[L_{1-\alpha}(x,n); U_{1-\alpha}(x,n)]$ are defined as $w_L = \frac{x}{n} - L_{1-\alpha}(x,n)$ and $w_U = U_{1-\alpha}(x,n) - \frac{x}{n}$, the distances between the point estimate and the CI bounds. The local average half-widths are defined as the expected half-widths in experiments with a random theoretical proportion $P$, as described in the previous section, and are denoted $w_L'' = E[W_L]$ and $w_U'' = E[W_U]$. Expected half-widths will be analyzed as functions of $\lambda$, the expected number of successes.

### 2.2.4. Relative local average half-widths

Local average half-widths are highly dependent on the number of successes. This makes CI widths difficult to compare on graphical figures. In order to make comparisons easier, relative local average half-widths have been defined as the ratio between the local average half-width of the CI estimator and the local average half-width of a reference CI estimator for the same actual proportion. The Clopper-Pearson mid-P estimator has been selected as reference for its good statistical properties.

### 2.2.5. Other evaluation criteria

The following few desirable properties of CIs have been assessed (shown in Appendix 3):

1) Consistency of CI inference and p-values of hypothesis tests [10];
2) Generalizability to bivariate and multivariate models;
3) Theoretical simplicity and the existence of an analytical solution. This motivated Agresti and Coull when they defined their CI [1,11];
4) Equivariance: The consistency of the CI of successes with the CI of failures [12–14];
5) Strict monotonicity of CI bounds along $x$, $n$ and $\alpha$ [15];
6) Deterministic procedure. Randomized CIs, based on a computer-generated random number, produce different CIs for the exact same data set. Conditional risks are smoothed by randomization, but the practical use of these CI requires a rigorous analysis difficult to apply in practice [16].

## 2.3. Graphical representations

Most interpretations will be graphical. Coverage bias will be expressed by local average errors. In order to help identify acceptable and unacceptable bias, reference lines will be drawn on graphical figures at thresholds $0.025 \times 1.50 = 0.0375$ and $0.025/1.50 = 0.016667$ for nominal one-sided risks at 0.025. These thresholds seemed relevant to article authors' but are arbitrary.

## 2.4. Validation

All analyzed CI have been implemented by the first author of this article with the R statistical software except the Blyth-Still-Casella CI, the C++ implementation by Keith Winstein (commit 850c75a35f816aa22fd6050453f1b7df2c5773c6) [17]. It has been verified to be consistent with StatXact on a few examples, in order to avoid software bugs. When an implementation was available in an R package (Hmisc, binom, exactci, DescTools, PropCIs), consistency has been verified.

In order to avoid transcription errors in formulas shown in tables of this article, the algorithms have been implemented again, from these formulas, two month later, by the same author. Once the second implementation was completed, an automatic consistency check was performed. A few transcription errors were fixed thanks to this second check. This double implementation has not been performed for the most complex CIs (Wang [18], Schilling-Doi [19] and Blyth-Still-Casella [20]) but consistency with author's original script had been verified. A third implementation has been performed by the same author, before manuscript finalization, months later, for the nine CIs presented in Table 1.

# 3. Results

## 3.1. Flow chart of CI estimators

From the systematic review 64 CI estimators were found. Four additional custom CI estimators were built by fixing problems in existing estimators (Appendix 1, two Pan 2002 modifications, a $BC_a$ bootstrap modification and a modification of the likelihood ratio interval), leading to a total of 68 CI estimators. Thirteen estimators were excluded. Approximations of Molenaar [21], Pratt [22], Blyth's equation C improving the Molenaar approximation [22] and the Chen CI [23] were all accurate approximations of the Clopper-Pearson CI [3] and so were considered redundant. The estimator of Zhou [24] and the average randomized inference model (ARIM) estimator of Lu [5] were redundant with Bartlett's Arc-Sine estimator while the inference model (IM) estimator of Lu [5] was redundant with the Clopper-Pearson mid-P estimator. The estimators of Crow was redundant with Blaker's [25]. Probit and logit transformations for Wald's CI were redundant. Estimators of Stevens [16], Geyer [26] and Zieliński [27] were randomized or fuzzy. The Wang 2017 [28] estimator is deterministic but relies on ranks of successes in addition to the sufficient statistics (number of successes and number of trials); it has not been analyzed.

Eventually, 55 CI estimators have been extensively analyzed. Nine are presented in this article (see Table 1) and the 46 others (including boostrap CIs and closed-form skewness corrected CIs) are presented in Appendix 1.

## 3.2. Definition of CI estimators

The lower bounds $L_{1-\alpha}(x,n)$ of estimators are described in Table 1. Upper bounds of these nine CI estimators can be computed by equivariance [12], i.e. $U_{1-\alpha}(x,n) = 1 - L_{1-\alpha}(n-x,n)$. In all cases, bounds outside the $[0,1]$ interval were set to the nearest valid proportion, 0 or 1. For example, when Wald's CI had a negative lower boundary, it was set at zero. These nine CI estimators have been selected because they are widely used (Wald, Clopper-Pearson), have been recommended (modified Wilson, modified equal-tailed Jeffreys [2]), are special cases of widely used general estimators or models (Likelihood ratio, Wald logit), have good statistical properties (Bartlett Arc-Sine, Clopper-Pearson mid-P) or illustrate a specific problem (Blaker).

Wilson's CI can be obtained by inversion of chi-square tests without transformations. It has an analytical solution (root of a 2$^{nd}$ degree equation). The modified Wilson CI, based on a Poisson approximation for small values of $k$, was described by Brown, Cai and DasGupta, (see page 112) [2] but these authors did not specify the threshold $x^*$ for $n > 100$. Since they did not analyze the behavior of the CI for $n > 100$, they did not make a recommendation (personal communication of the first author). The already sufficient convergence to the Poisson distribution made us retain the threshold $x^* = 3$ for $n > 100$.

Bartlett's Arc-Sine CI is a normal-approximation CI with a variance-stabilizing transformation. Variance stabilization has been improved from the standard Arc-Sine CI transformation, adding 0.5 success and 0.5 failures.

The Wald logit CI is the normal-approximation asymptotic CI that statistical software typically computes for intercept-only logistic regressions. It is indefinite for $x = 0$ and $x = n$. We supplement its definition by the Clopper-Pearson CI for $x = 0$ and $x = n$.

The likelihood ratio CI is defined by inverting a chi-square test on the deviance function. This CI is well defined even for $x = 0$ and $x = n$, but in order to compare its performances with those of the Wald logit CI, we applied the same Clopper-Pearson substitution for $x = 0$ and $x = n$. The unmodified likelihood ratio CI is presented in, Figures A.8 and A.10 of Appendix 1.

Jeffreys CI is an equal-tailed bayesian credible CI with the non-informative Jeffreys prior $Beta\left(\frac{1}{2};\frac{1}{2}\right)$. Brown *et al* [2] proposed a slight modification to improve its frequentist properties. The lower bound of the CI is set at zero when the number of successes ($x$) is zero or one and the upper bound is set to the same exact upper Poisson boundary as that of Wilson's modified CI when the number of successes is zero

In table 1, we defined three exact binomial CIs that can be constructed by test inversion: (i) Clopper-Pearson, (ii) Clopper-Pearson mid-P and (iii) Blaker. The first is constructed from a one-sided exact binomial test with non-strict inequality: $\Pr(X \geq x)$. The second is similar to the first but uses a "half strict" inequality: $\Pr(X \geq x) + \frac{1}{2}\Pr(X = x)$. Blaker's CI is based on a two-sided test. Blaker's test *P*-value, for $x > 0.50$ is equal to $\Pr(X \geq x) + \Pr(X \leq y)$ where $y$ is the largest number such as $\Pr(X \leq y) \leq \Pr(Y \leq y)$. For $x < 0.50$, the CI can be defined by equivariance.

## 3.3. Results common to all CI estimators

Local average errors (Figure 1) are smoothed in comparison to conditional errors (Figure 2) having large amplitude oscillations. For a constant expected number of successes $\lambda$, coverage bias is higher in absolute value for larger values of $n$. The results for $n = 2048$ are very close to those for asymptotic Poisson CIs (see Figure 2 and Figures A.4 and A.5).

Local average interval half-widths mirror local average errors (Figure 3). Where a CI is shorter than another, it has a higher local average error and where it is larger, it has lower local average error.

## 3.4. Specific CI results

Wald's CI has a high two-sided bias, even for an expected number of successes equal to 32, and is a very unbalanced unequal-tailed CI (see Figures 1-2). The right local average error $\alpha'_U$ tends to 1 when the expected true proportion $p_0$ tends to zero because Wald's CI width is zero when the number of successes $x$ is null.

Modified Wilson's, modified Wald logit and Blaker's CIs have lower absolute bias but are not equal-tailed either. The biases of these three CIs estimators have an opposite sign to Wald's CI estimator bias, while the modified likelihood ratio CI has a small bias in the same direction as Wald's CI. The modified Wald logit CI has a lower bound local average error spike ($\alpha''_L$) equal to 0.097 for $n = 2\,048$ and an expected number of successes $np_0 = \lambda = 0.11$.

Blaker's and Clopper-Pearson CIs are both conservative, slightly less so for Blaker's CI (Figures 1-2). For proportions close to zero, Blaker's CI conditional right $\alpha'_U$ error (Figure 2) is very close to Clopper-Pearson $\alpha'_U$ error but Blaker's $\alpha'_L$ errors can get much higher with one-sided conditional error oscillations up to 0.05 while Clopper-Pearson's CI one-sided conditional error oscillations never exceed 0.025.

Local average errors (Figure 1) with Bartlett's Arc-Sine CI, modified equal-tailed Jeffreys and Clopper-Pearson mid-P CIs are close to each other. The modified equal-tailed Jeffreys CI, for proportions close to zero, has a larger right local average half-width than Bartlett's Arc-Sine and Clopper-Pearson mid-P CIs (Figure 3) leading to a more conservative upper boundary for an expected number of successes close to 4 (Figure 1). Bartlett's Arc-Sine CI local average errors get closer to nominal than Anscombe and Freeman-Tukey CIs (see Figure A.7).

The modified likelihood ratio CI has a mild one-sided local average bias, lower than the Wald, modified Wald logit and modified Wilson CIs (Figure 1). The unmodified approximate likelihood ratio CI has a high one-sided error spike for an expected number of successes close to 2.3 (Figure A.8).

Percentile and basic bootstrap CIs (Figure A.8) have high local average biases. The basic bootstrap CI has higher biases than Wald's CI while the percentile bootstrap CI is slightly less biased than Wald's. Unmodified $BC_a$ bootstrap is highly biased and is not equivariant; a modification of equation 3.2 of Efron [29] taking in account the discreteness of the bootstrap distribution, provides equivariance and reduces the bias (Appendix 1). Modified $BC_a$ bootstrap, smoothed $BC_a$ bootstrap and studentized $BC_a$ bootstrap CIs are all conservative in terms of local average error. Due to division by zero errors, the studentized bootstrap CI cannot be computed for $\min(x, n - x) \leq 4$ and has been replaced by the Clopper-Pearson CI.

Not all two-sided unequal-tailed CIs have the same imbalance of errors. Extreme examples are shown in Appendix 1. The Pan 2002 Wald $t$ CI (Table A.1 and Figure A.7) for sample size $n = 2048$ and expected number of successes $np_0 = 8.01$ has $\alpha_L'' = 0.005$ and $\alpha_U'' = 0.045$ while the Rubin logit CI (Table A.3 and Figure A.8) for the same experiment has $\alpha_L'' = 0.045$ and $\alpha_U'' = 0.003$.

Other evaluation criteria are shown in Appendix 3. All the nine CIs shown in Table 1 are equivariant and deterministic. Generalizability to bivariate and multivariate models applies to the Wald, Wald logit, likelihood ratio, Clopper-Pearson and Clopper-Pearson mid-P CIs. Strict monotonicity of CI bounds along $x$, $n$ and $\alpha$ is guaranteed for all the nine CIs except Blaker's CI, the modified equal-tailed Jeffreys CIs and Wald's CI. No analytical solution is known for the likelihood ratio, Blaker and Clopper-Pearson mid-P CIs while the six other CIs have one.

## 3.5. Continuity correction

Continuity corrections make CIs more conservative on both sides (Appendix 1, Figures A.7-9). Regarding Wilson's and Wald's CIs make them less liberal on one side (respectively left and right side) and more conservative on the other (respectively right and left side).

## 3.6. Sensitivity analyses

For a typical odds ratio between the actual proportions of two experiments $OR_S = 1.20$, graphically, local average error oscillations can be seen for an expected number of successes below 2 (Figure 1). For an actual proportion with less random fluctuation, $OR_S = 1.05$, large amplitude oscillations are graphically visible when the expected number of successes is below 8 (Appendix 1, Figure A.2), with a maximum local average one-sided error equal to 0.0381 for the Clopper-Pearson mid-P CI.

The 90% CI absolute coverage biases are larger than 95% CI biases (as seen from comparing Figure 1 with Figure A.3) but relative biases, i.e., ratios of actual to nominal errors, are smaller.

The random sample average risks $\alpha'''$ of a constant proportion $p$ with random sample size $N$ are close to local average $\alpha''$ errors for random proportion $P$ with a constant sample size $n$ (see Figure 1 and A.6).

## 3.7. Validity conditions of Wald's CI

Validity conditions of Wald's CI are assessed in Appendix 1 (Tables A.12-14). The simple condition $\min(x, n - x) > 40$ is enough to control the one-sided local average error for a 95% CI albeit not perfectly as the actual one-sided error may be up to 1.5 times higher (that is 0.0375) than the nominal error (0.025).

# 4. Discussion

## *4.1. Summary of main findings*

Wald's CI has much higher biases than other CI estimators and is unequal-tailed. The modified Wilson CI and modified Wald logit CIs have lower local average biases but are unequal-tailed too. Wald's logit CI has higher biases than the likelihood ratio CI.

Bartlett's Arc-Sine, Clopper-Pearson mid-P and modified Jeffreys equal-tailed CIs are close to each other in terms of the local average error control and expected CI half-widths. They properly control the local average errors and are equal-tailed.

## *4.2. Recommendations*

Statistical software (e.g. R, SAS, SPSS) provides two CI estimators for generalized linear models coefficients: likelihood ratio CI and Wald's CI. We showed lower coverage bias for the likelihood ratio CI than for Wald's CI. This suggests that likelihood ratio CIs may be better for logistic regressions. This is consistent with Agresti's [30] recommendation suggesting the use of the likelihood ratio CI when Wald's CI is discordant with it.

In our opinion, Bartlett's Arc-Sine, Clopper-Pearson mid-P and modified Jeffreys equal-tailed CIs have the best statistical properties and are practically equivalent. The modified Jeffreys CI is based on *ad hoc* modifications [2] and has a lower bound that is not strictly monotone with the number of successes since the lower bound is equal to zero for a number of successes equal to zero or one. Therefore, we cannot recommend the equal-tailed Jeffreys CI. The Clopper-Pearson's mid-P CI has been generalized to logistic regressions by Hirji [31], making it theoretically more attractive than Bartlett's Arc-Sine CI. Therefore, we recommend using the Clopper-Pearson's mid-P CI in almost all scenarios.

When comparing an observed proportion to a theoretical proportion under highly controlled experimental conditions or in a strongly regulated domain such as clinical trials, we recommend the Clopper-Pearson CI. Indeed, the theoretical proportion $p$ is not variable anymore, as it is defined in a protocol, and the sample size may be quite controlled; oscillations are not smoothed anymore and the Clopper-Pearson CI (strictly conservative) may be safer than the Clopper-Pearson mid-P CI (control local average errors but not conditional errors).

## *4.3. Originality of this work*

Newcombe distinguished mesial (the CI bound nearest to 0.50) and distal (the CI bound nearest to 0 or 1) one-sided errors [7] but averaged them over the whole $]0,1[$ interval assuming a uniformly distributed random proportion. Agresti and Coull [1] analyzed a random proportion following a beta distribution with 0.10 expectancy, which is similar to the local average error for an expected true proportion $p_0 = 0.10$. However, they did not analyze other theoretical random proportions or one-sided errors. To our knowledge, the influence of the sample size for a fixed expected number of successes has not been graphically presented in any systematic review of binomial proportion CIs. These new evaluation criteria, including one-sided errors of two-sided CI, make the originality of this work.

## 4.4. Validity conditions of Wald's CI

Wald's CI has low coverage even for quite high number of success and failures, leading some authors to recommend against teaching this CI in elementary courses, favoring Agresti-Coull or Wilson's CI [1,2,11]. Recent textbooks such as Fritz and Berger in 2015 [32] may mention that Wald's CI is biased, citing Agresti and Coull, but then give the old validity condition $np, n(1-p) \geq 5$. The condition $\min(x, n - x) > 40$ controls the one-sided local average errors for a 95% CI. It may be taught in place of $\min(x, n - x) \geq 5 \text{ or } 10$.

## 4.5. Poisson distribution

As the binomial distribution is asymptotically equivalent to the Poisson distribution when the sample size is infinite and the expected number of successes is held constant, results for binomial CIs for a large sample size ($n = 2048$) can be extrapolated to the Poisson CIs. Convergence to the Poisson distribution has been shown in A.4 and A.5.

## 4.6. Bootstrap

According to Carpenter [33], under usual conditions, the theoretical convergence rate of bootstrap 'normal', studentized and percentile is of the order $O\left(\frac{1}{\sqrt{n}}\right)$ whereas the theoretical convergence rate of the studentized and $BC_a$ bootstrap is of the order $O\left(\frac{1}{n}\right)$. The 'normal' bootstrap is equivalent to Wald's CI. The asymptotical convergence rate of these CIs is reflected in the very high biases of the percentile and basic bootstrap CIs and much lower biases of the modified $BC_a$ and studentized bootstrap CIs. No bootstrap CI controls the nominal error as well as the Clopper-Pearson mid-P CI.

## 4.7. Implementations

We recommend using the Clopper-Pearson mid-P CI. Its computer implementation is available in the exactci package for the R statistical software and in SAS version 9.4 through the MIDP option of the EXACT statement of PROC FREQ. Macros or programs for SPSS, Stata, SAS (for older version), Python, Minitab, MYSTAT/SYSTAT, Microsoft Excel, LibreOffice and nine other software and platforms are given in Appendix 2. They are distributed under the free software Creative Commons CC0 license terms. Online implementation of the procedure is available [34,35]. The R source code used to calculate confidence intervals, figures and tables of this document is distributed under the CC0 license in Appendix 4.

# 5. Practical example

Dellas *et al* [36] published a prognostic study in a population of hemodynamically stable patients with acute symptomatic pulmonary embolism. The objective was to predict complicated courses within 30 days (death, catecholamine administration, mechanical ventilation and resuscitation) from clinical parameters, Heart-type Fatty Acid Binding Protein (H-FABP) and multidetector computed tomography (MDCT). The simplified Pulmonary Embolism Severity Index (sPESI) was used to assess the clinical risk.

Complication risks were reported by frequency and 95% CI of proportions. Several frequencies were low: one adverse outcome for 225 patients was reported in the group defined by H-FABP ≤ 6 ng/mL and sPESI = 0. Two adverse outcomes for 46 patients were reported in the group defined by H-FABP > 6 ng/mL and sPESI = 0. The reported CIs were 1/225 (0.4%, 95% CI: 0 to 1.3%) and 2/46 (4.3%, 95% CI: 0 to 10.9%). The CI estimator was not reported in the methods section. Different estimators yield quite different results, as shown in Table 2.

Dellas *et al* most probably used the percentile bootstrap estimator, as could be verified from the many low frequency proportions in the article with numerators ranging from 1 to 6. For the proportion 1 / 225, the upper boundary of the proportion ranged from 0.9% to 3.1%, depending on the estimator and the lower boundary could be negative, zero or positive.

It's not easy to estimate the actual confidence level Della *et al* had, since the actual proportion is unknown. It can be assumed, from the Clopper-Pearson mid-P CI that the actual proportion is less than 2.18% (upper boundary of the 95% CI). We assume that the authors would have noticed that the percentile bootstrap yields a [0 ; 0] CI when the observed proportion is zero and would have used the Clopper-Pearson 95% CI (classical exact CI) in place. In that conservative scenario, the actual local average lower bound and upper bound risks would have been respectively 0.93% and 7.4%. Therefore, the CI risks are unbalanced and far from the 2.5% nominal risk. The classical two-sided conditional coverage for this CI, for this theoretical proportion (2.18%) is equal to 94.6%, with actual risks of 1.1% of overestimation ($\alpha'_L$ risk) and 4.3% of underestimation ($\alpha'_U$ risk). This would appear to be a small bias.

This example illustrates that the classical assessment of estimator bias by two-sided conditional coverage may provide results quite different from one-sided local average errors. In our opinion, the one-sided local average errors are more relevant as they take in account the variability of the number of subjects and/or actual proportion from one experience to another and allow interpretation of each CI boundary separately. This also shows that while bootstrap methods are asymptotically unbiased, on small or unbalanced samples they may be unreliable. This also illustrates the need of specifying the actual statistical method used in scientific articles.

# 6. Conclusion

The binomial proportion CI problem may seem trivial, but some aspects of the problem may be missed, such as tails equality or local average error control.

In this light, we assessed 55 CI estimators and give the following recommendation: use the Clopper-Pearson mid-P CI in all scenarios but the comparison of an observed proportion to a theoretical proportion in a strongly controlled or heavily regulated experimental environment, such as a clinical trial; in that case, use the Clopper-Pearson CI.

# 8. Table and figures

Table 1: definition of lower bounds of the confidence intervals, upper bounds being defined by equivariance $U_{1-\alpha}(x,n) = 1 - L_{1-\alpha}(n-x,n)$ according to sample size $n$ and number of sucesses $x$

| Name | Lower bound $L_{1-\alpha}(x,n)$ |
|---|---|
| Wald[a] | $\max\left(0, \dfrac{x}{n} - \kappa\sqrt{\dfrac{x(n-x)}{n^3}}\right)$ |
| [2] Modified Wilson[a,b] | $\begin{cases} \dfrac{1}{2n}\chi^2_{\alpha,2x} \text{ if } 1 \leq x \leq x^* \\ \dfrac{x + \dfrac{\kappa^2}{2} - \kappa\sqrt{\dfrac{x(n-x)}{n} + \dfrac{\kappa^2}{4}}}{n + \kappa^2} \text{ otherwise} \end{cases}$ <br> where $x^* = 2$ for $n \leq 50$ and $x^* = 3$ for $n > 50$ |
| [37,38] Bartlett Arc-sine[a] | $\sin^2\left(\max\left(0, \operatorname{asin}\left(\sqrt{\dfrac{x + \dfrac{1}{2}}{n+1}}\right) - \dfrac{\kappa}{2\sqrt{n + \dfrac{1}{2}}}\right)\right)$ |
| [2] Modified Wald logit[a,c] | $\begin{cases} \operatorname{logitinv}\left(\log\left(\dfrac{x}{n-x}\right) - \kappa\sqrt{\dfrac{n}{x(n-x)}}\right) \text{ if } 0 < x < n \\ \sqrt[n]{\alpha/2} \text{ if } x = n \\ 0 \text{ if } x = 0 \end{cases}$ |
| [6] Modified likelihood ratio[a] | $\begin{cases} \inf\left\{q \,\Big|\, \log\left(\left(\dfrac{x}{nq}\right)^x \left(\dfrac{n-x}{n(1-q)}\right)^{n-x}\right) \leq \dfrac{1}{2}\kappa^2\right\} \text{ if } 0 < x < n \\ \sqrt[n]{\alpha/2} \text{ if } x = n \\ 0 \text{ if } x = 0 \end{cases}$ |
| [2] Modified equal-tailed Jeffreys[d] | $\begin{cases} \beta iCDF(\alpha/2; x + 1/2, n - x + 1/2) \text{ if } 2 \leq x < n \\ \sqrt[n]{\alpha/2} \text{ if } x = n \\ 0 \text{ if } x \leq 1 \end{cases}$ |
| [39] Blaker | $\inf\{q | \operatorname{bpval}(q, x, n) > \alpha\}$ <br> where <br> $\operatorname{bpval}(p, x, n)$ <br> $= \begin{cases} \Pr(X \leq x \text{ or } X \geq \inf\{x' | \Pr(Y \geq x') \leq \Pr(Y \leq x)\}) \text{ if } p \geq \dfrac{x}{n} \\ \operatorname{bpval}(1-p, n-x, n) \text{ if } p < \dfrac{x}{n} \end{cases}$ <br> Where $X \sim \mathcal{B}(n,p)$ and $Y \sim \mathcal{B}(n,p)$ |
| [2,3] Clopper-Pearson[d] | $\beta iCDF\left(\dfrac{\alpha}{2}; x, n - x + 1\right)$ |
| [40,41] Clopper-Pearson mid-P | $\inf\left\{q \,\Big|\, \Pr(X \geq x) - \dfrac{1}{2}\Pr(X = x) > \dfrac{\alpha}{2} \text{ where } X \sim \mathcal{B}(n,q)\right\}$ |

[a]We denote by $\kappa = z_{1-\alpha/2}$ the quantile $1 - \alpha/2$ of the normal distribution $\mathcal{N}(0,1)$

[b]$\chi^2_{q,df}$ is the $q$ quantile of the $\chi^2$ distribution with $df$ degrees of freedom

[c] The reciprocal of the logistic transformation is defined by $\text{logitinv}(t) = \frac{\exp(t)}{1+\exp(t)}$

[d] $\beta iCDF(q; \alpha, \beta)$ is the $q$ th quantile of the beta distribution whose shape parameters are $\alpha$ and $\beta$

**Table 2: different CI estimators applied to the proportions defined in the article of Dellas *et al*.**

|  | Observed proportion | |
| --- | --- | --- |
|  | 1 / 225 (0.4%) | 2 / 46 (4.3%) |
| **Estimator** | | |
| Percentile bootstrap | 0 to 1.3% | 0 to 10.9% |
| Basic bootstrap | −0.4% to 0.9% | −2.2% to 8.7% |
| Wald CI | −0.4% to 1.3% | −1.5% to 10.2% |
| Clopper-Pearson | 0.01% to 2.4% | 0.5% to 14.8% |
| Clopper-Pearson mid-P | 0.02% to 2.2% | 0.7% to 13.6% |
| Wilson | 0.08% to 2.5% | 1.2% to 14.5% |
| Wald logit | 0.06% to 3.1% | 1.1% to 15.8% |
| Likelihood ratio | 0.03% to 1.9% | 0.7% to 12.8% |

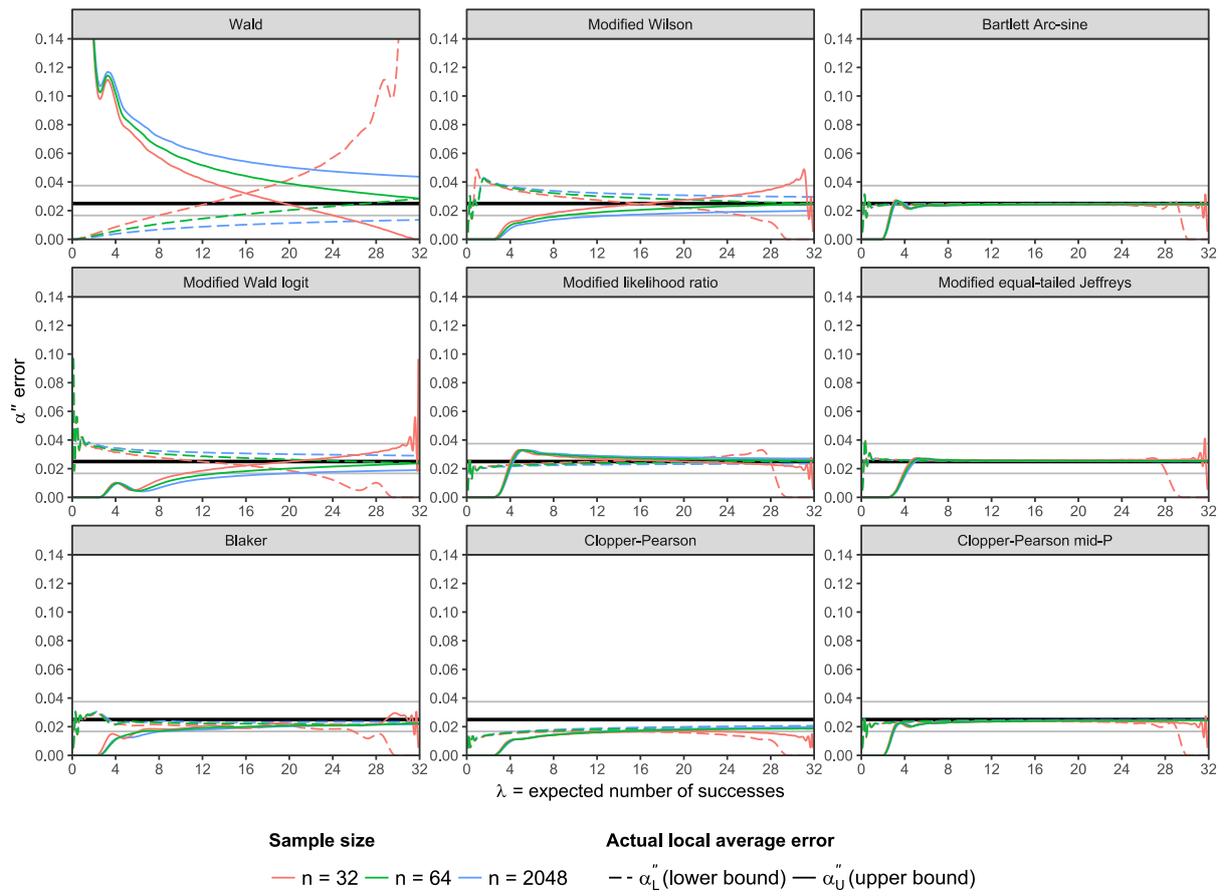

Figure 1: one-sided local average errors of nine 95% confidence interval estimators according to different sample sizes (red for $n = 32$, green for $n = 64$ and blue for $n = 2048$), with random actual $P$ proportion following a logit-normal distribution with typical odds ratio of the actual proportion between two experiments equal to $OR_S = 1.20$. The abscissa is the expected number of successes $np_0$ and the ordinate is the probability that the lower bound of the confidence interval is greater than the true proportion $p$ (left local average error: dashed lines) or the probability that the upper bound of the confidence interval is lower than the true proportion $p$ (right local average error: solid line).

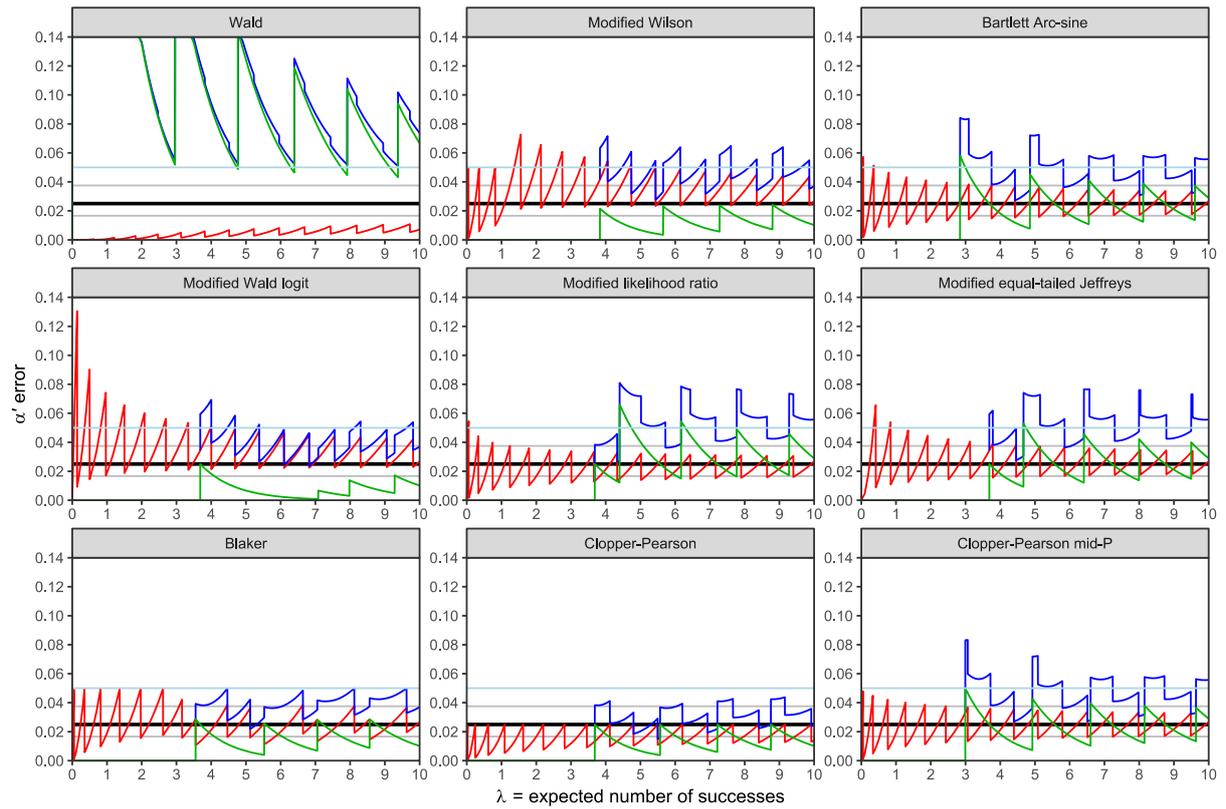

**Figure 2:** one-sided and two-sided conditional errors of nine 95% confidence interval estimators for a sample of size $n = 2048$ and a constant theoretical $p$ proportion. The abscissa is the expected number of successes $np$ and the ordinate is the risk that the lower bound of the confidence interval is greater than the true proportion $p$ (left conditional error: red), the risk that the upper bound of the confidence interval is lower than the true proportion $p$ (right conditional error: green) or the risk that the confidence interval does not contain the true proportion $p$ (two-sided risk: blue).

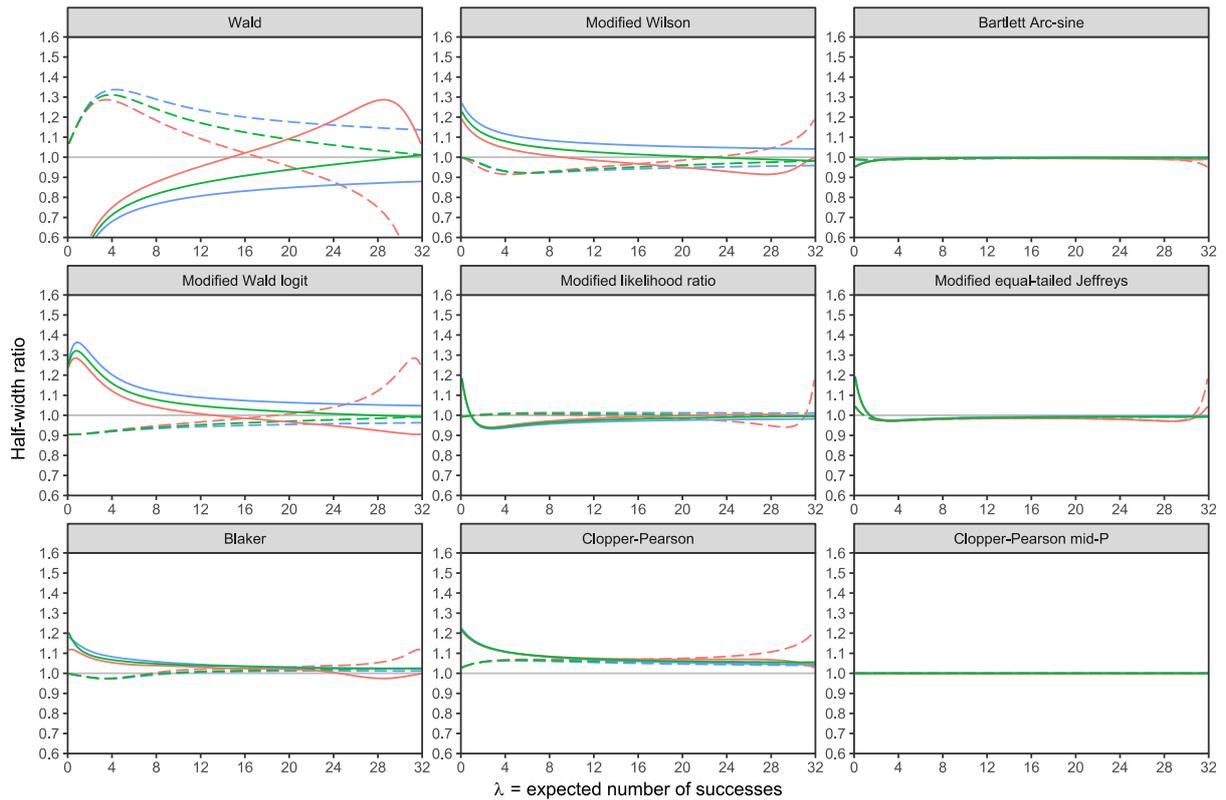

Figure 3: relative local average half-widths of nine 95% confidence interval estimators for a sample of size $n = 2048$ with a random actual $P$ proportion following a logit-normal distribution with a typical odds ratio of the actual proportion between two experiments equal to $OR_S = 1.20$. The relative half-width is the local average half-width of one of the nine intervals divided by the local average half-width of the Clopper-Pearson mid-P interval for the same $x$, $n$ and $p_0$ (expected $P$) parameters. The abscissa is the expected number of successes $np$ and the ordinate is the left relative local average half-width (dashed lines) or the right relative local average half-width (solid lines).

# 9.  Appendices

All appendices are available on the Open Science Framework (https://osf.io/gqyem/)